\documentclass[apj,letterpaper]{emulateapj}

\usepackage{apjfonts}
\usepackage{amsmath}
\usepackage{aas_macros}
\usepackage{graphicx}
\usepackage{natbib}
\usepackage{xspace}
\usepackage{longtable}
\usepackage{multirow}

\newcommand{\hinvMsun}{\ensuremath{h^{-1}{{\rm M}_\odot}}\xspace}
\newcommand{\Mvir}{\ensuremath{\rm M_{vir}}\xspace}
\newcommand{\Rvir}{\ensuremath{\rm R_{vir}}\xspace}
\newcommand{\Vvir}{\ensuremath{\rm V_{vir}}\xspace}
\newcommand{\rsep}{\ensuremath{\rm r_{sep}}\xspace}
\newcommand{\Rsep}{\ensuremath{\rm R_{sep}}\xspace}
\newcommand{\vsep}{\ensuremath{\rm v_{sep}}\xspace}
\newcommand{\vinfall}{\ensuremath{\rm v_{infall}}\xspace}

\newcommand{\Rhalf}{\ensuremath{\rm R_{half}}\xspace}
\newcommand{\Vsep}{{\rm V_{sep}}\xspace}
\newcommand{\rhoc}{\ensuremath{\rho_{\rm crit}}\xspace}

\newcommand{\SUB}{{\small{SUBFIND}}\xspace}
\newcommand{\Ob}{\ensuremath{\Omega_{\rm b}}\xspace}
\newcommand{\Ol}{\ensuremath{\Omega_\Lambda}\xspace}
\newcommand{\Odm}{\ensuremath{\Omega_{\rm dm}}\xspace}
\newcommand{\sig}{\ensuremath{\sigma_8}}
\newcommand{\FOF}{{\small{FOF}}\xspace}

\newcommand{\dphi}{\ensuremath{\Delta E/E}\xspace}

\newcommand{\kms}{{\rm km/s}}

\newcommand{\kpch}{\ensuremath{h^{-1} {\rm kpc}}\xspace}
\newcommand{\mpch}{\ensuremath{h^{-1} {\rm Mpc}}\xspace}
\newcommand{\LCDM}{\ensuremath{\Lambda {\rm CDM}}\xspace}
\newcommand{\phno}{\phantom{0}}%
\newcommand{\phnt}{\phantom{00}}%

\newcommand{\massratioindex}{-1.1}
\newcommand{\ml}{\ensuremath{\Upsilon_*}}

\newcommand{\var}[2]{\xspace{\ensuremath{{#1}_{#2}}\xspace}}

\begin{document}

\title{A First Look at Galaxy Flyby Interactions. II. Do Flybys matter?}

\author{Manodeep Sinha\altaffilmark{1}, and Kelly Holley-Bockelmann\altaffilmark{2}}
\affiliation{Department of Physics and Astronomy, Vanderbilt University, Nashville, TN, 37235}
\altaffiltext{1}{manodeep.sinha@vanderbilt.edu}
\altaffiltext{2}{k.holley@vanderbilt.edu}

\begin{abstract}
In the second paper of this series, we present results from cosmological simulations on
the demographics of flyby interactions to gauge their potential impact on galaxy evolution. In a previous paper, we demonstrated that
flybys -- an interaction where two independent halos inter-penetrate but detach at a later time and do not merge -- occur
much more frequently than previously believed. In particular, we found that the frequency of flybys increases at low redshift
and is comparable to or even greater than the frequency of mergers for halos $\gtrsim 10^{11}\, \hinvMsun$. 
In this paper, we classify flybys according to their orbits and the level of perturbation exacted on both the halos involved. 
We find that the majority of flybys penetrate deeper than $\sim \Rhalf$ of 
the primary and have an initial relative speed $\sim 1.6\times\Vvir$ of the primary.
The typical flyby mass-ratio is $\sim 0.1$ at high $z$ for all halos, while at low $z$,
massive primary halos undergo flybys with small secondary halos. 
We estimate the perturbation from the flyby on both the primary and the secondary
and find that a typical flyby is mostly non-perturbative for the primary halo. 
However, since a massive primary experiences so many flybys at any given time, they
are nearly continually a victim of a perturbative event.  In particular, we 
find flybys that cause $\sim 1\%$ change in the binding energy of a primary halo occurs 
$\gtrsim 1 $ Gyr$^{-1}$ for halos $> 10^{10}\,\hinvMsun$ for $z \lesssim 4$. Secondary halos,
on the other hand, are highly perturbed by the typical encounter,
experiencing a change in binding energy of nearly order unity. Our results imply
that flybys can drive a significant part of galaxy transformation at moderate to lower redshifts ($z \lesssim 4$).
We touch on implications for observational surveys, mass-to-light ratios, and galaxy assembly bias. 

\end{abstract}


\maketitle

\section{INTRODUCTION}

Galaxy mergers drive galaxy evolution; they are a key mechanism by which galaxies grow and transform.
The merger process is usually accompanied by a strong gas inflow~\citep{BH91, MH94}
triggering a central starburst and perhaps an active galactic nucleus (hereafter, AGN).
Consequently, galaxy mergers have been studied quite extensively both
numerically~\citep[e.g.,][]{DMH99,RBCD06,CJ08,HCH08,SH09,SH10} and
observationally~\citep[e.g.,][]{S86, L03, VD05, LD08,TGSN08}.
Galaxy mergers are so successful in driving galaxy evolution simply because they
strongly perturb the potential. Of course, weaker perturbations can cause change, too;  it has
long been thought that even an orbiting low mass satellite like the Large Magellanic Cloud can
distort the underlying smooth galaxy potential enough to warp the Milky Way (MW) disk~\citep{WB06}.
However, one entire class of galaxy interactions capable of causing such perturbations -- flybys --  has been largely ignored.

Unlike galaxy mergers where two galaxies combine into
one remnant, flybys occur when two independent galaxy halos inter-penetrate but
detach at a later time; this can generate a rapid and large perturbation in each galaxy. We define
a flyby as a fast interaction between two independent galaxies each within their 
own separate dark matter halos.
Although some aspects of a galaxy flyby may be similar to the first wide pass of a galaxy merger, or to
galaxy harassment in a cluster, the rapid and transient nature of a flyby places it in a dynamical
class of its own.

One of the first questions to ask is {\it how common are galaxy flybys?}. Curiously, though, this is a difficult question to answer
simply because previous theoretical work has been set up to search for mergers; indeed both the extended Press-Schechter framework
hierarchically assembles a dark matter halo, and the merger trees gleaned from
cosmological Nbody simulations are designed to track halo growth in the same way -- solely through mergers and smooth accretion.

Previous numerical works have shown the existence of flybys, or `backsplash' galaxies, in galaxy clusters~\citep[e.g.,][]{GKG05,MSSS04} and
isolated galaxies~\citep{WKP08,KLKMYGH11}. Simulations predict that up to $\sim 60\%$ of the galaxies found between $1-2\,\Rvir$
could be `backsplash' galaxies~\citep{GKG05,P11,BMBF12}. Naturally, these galaxies are distinct from those that are infalling
for the first time. Backsplash galaxies are subject to `pre-processing' and usually exhibit HI deficiency
compared to infalling satellites. Candidate backsplash galaxies have been identified in galaxy clusters~\citep{MSSS04,MMR11,P11},
as well as the field galaxies in the Local Group~\citep{TJK12}. Another theoretical approach at finding flybys focused on identifying subhalo orbits that extend far
outside the primary virial radius~\citep{LNSJFH09,WJ09,BLW12}. \citet{LNSJFH09} looked at a suite of five isolated Milky-Way sized dark matter halos in a cosmological simulation 
and found that roughly half of the subhalos that were once associated with the host are now located outside the virial radius. Broadly speaking, flyby halos
constitute $\sim$ 50\% of all halos on scales of  $1-3 \,\Rvir$.

Recently, we developed and tested a method to identify both mergers and flybys between dark matter halos in
cosmological simulations~\citep{SH12}. We constructed a full `halo interaction network' that assesses the past dynamical
history of any given halo. With this new tool\footnote{available publicly at \url{https://bitbucket.org/manodeep/hinge}}, 
we made the first census of flybys, and we found that they are surprisingly common. In fact at $z<2$, flybys occur more often than
mergers for all halos $\gtrsim 10^{11}\hinvMsun$. 

We reiterate that majority of flybys are one-time events, not merely the first orbit of a merger -- here, the disks (if present)
will not physically overlap at any point; 
 they are also fundamentally different than galaxy harassment. More cosmological simulations are underway to build up a better
statistical sample and to expand the dynamic range in halo mass, but the emerging picture is that current studies focusing on solely merging galaxies
may completely miss this common type of galaxy interaction.

Now that we have evidence that flybys are rampant, we need to know if neglecting these flybys matters.
While dark matter halo flybys do not significantly affect the halo mass function, these fast, transient events may make their mark on
galaxy structure and kinematics.  Over the years, flybys have been invoked in a heuristic way to explain various morphological features and transformations, such as the 
evolution from spiral to S0 galaxies in group environments~\citep{BC11}, or the excitation of spiral arms in the galactic disk~\citep[e.g.,][]{TF06}.
Recently, flybys have been proposed as a way to provide the extra starbursts needed to 
explain the steep abundance gradient found in massive ellipticals~\citep{CM11}, because it is widely-held that these galaxies
are too $\alpha$-enhanced to be explained by mergers or galaxy harassment alone~\citep{NLOB05,TMBM05}. 
From the purview of linear perturbation theory, 
a flyby imparts an impulse to the galaxy~\citep{VW00}, and in principle this can exact a morphological and kinematic change
in both the `victim' and the `perturber' halo. In fact, high resolution isolated simulations have borne this out, showing that flybys can induce 
a strong and long-lived bar~\citep{LHS14}, can spin up the dark matter halo~\citep{LHS14}, and can generate an S-shaped warp as well~\citep{KPKAY14}.

In this paper, we aim to quantify the flyby perturbation and the frequency of disruptive flybys as a function of 
halo mass and redshift. In Section~\ref{sec:methods} we briefly describe the simulation details and the interaction 
network (see \citet{SH12} for further details), in Section~\ref{sec:results} we show the results and Sections~\ref{sec:discussion}
and \ref{sec:conclusion} we discuss the implications of our results and future steps. In addition, we outline the details
of the analytic approximation used to estimate the perturbation on the halo in Appendix~\ref{appendix:perturbations}.

\section{METHODS}\label{sec:methods}

We use a high-resolution dark-matter simulation with $1024^3$ particles in a 50 \mpch box 
with $\Ob = 0.044,\, \Odm = 0.214, \, \Ol = 0.742,\, \sig = 0.796, h = 0.719, n_s = 0.96 $, 
consistent with {\small WMAP-5} cosmology~\citep{K08}, where the symbols have their 
usual meaning.  The initial particle distribution is obtained from a Zel'dovich linear 
approximation at a starting redshift of $z=249$ and is evolved using the adaptive tree-code, 
{\small GADGET-2}~\citep{SYW01,S05}. The dark matter particles have a fixed gravitational 
softening length of $2.4$ co-moving \kpch at all epochs. 
We store 105 snapshots spaced logarithmically in scale-factor, $a = 1/(1+z)$, from $z=20$ to $0$. This translates
into a timing resolution $\lesssim 50$ Myr for $z\gtrsim 3$ and $\sim 150$ Myr for $z \lesssim 3$; in SH12, we
showed that a high snapshot cadence was essential in fully detecting the flybys in the volume.
Since the fundamental mode goes non-linear at $z=0$, we only present results up to $z=1$ where the 50 \mpch
box is still a representative cosmological volume. 

To identify halos, we use a canonical linking length $b=0.2$ ($\sim$ 10 \kpch) to find the Friend-of-Friends (\FOF) halos~\citep{DEFW85}; subhalos,
with a minimum of 20 particles ($\sim 10^8 \hinvMsun$), are identified using the \SUB algorithm \citep{SWTK01}. 
The \SUB algorithm identifies subhalos as bound structures around a density maxima; 
the background \FOF obtained at the end is comprised of particles that are bound in the general potential of the \FOF, but not bound to any
subhalo. For the remainder of this paper, all references to the \FOF halo will mean this bound background halo. For details on the \SUB
algorithm, we refer the reader to \citet{SWTK01}. We used an overdensity, $\rho_\Delta = 200\, \rhoc(z)$ to compute the virial 
radius of halos:
\begin{equation}
\Rvir = \left(\dfrac{3\Mvir}{4 \pi \rho_\Delta }\right)^{1/3}.
\label{eqn:rvir}
\end{equation}
Here, we used the bound mass of the halo reported by \SUB as $\Mvir$. 
Another commonly used overdensity is, $\rho_{\Delta, {\rm vir}} \approx (18 \pi^2 + 82x - 39x^2)\,\times\,\rhoc(z) $\citep{BN98}, where
$x = \Omega_{m}(z) - 1$. For \LCDM cosmology at $z=0$, $\rho_{\Delta ,{\rm vir}} \sim 96\, \rhoc$. For comparison, the virial radii obtained
at $z \sim 0$ by using $\rho_{\Delta, {\rm vir}}$ would be roughly $(200/96)^{1/3} \sim 1.27$ times larger than our adopted definition. 
At higher redshifts, $\rhoc \approx \rho_{\rm matter}$, and there is virtually no difference between our adopted $\rho_\Delta$ and 
$\rho_{\Delta, {\rm vir}}$. 

\subsection{Characterizing flyby orbits}

Our next step is to identify all interactions between halos of any type in the simulation. 
This done with our halo interaction network algorithm described in SH12. For every flyby we record the initial separation, $\rsep$,  
a minimum separation (impact parameter), $b$, 
between the halo centers during the encounter and the initial relative velocity between the halo centers, $\vsep$,
at the beginning of the encounter. These three quantities, along with the two halo masses and the initial redshift
are sufficient to crudely specify the interaction. In other words, idealized galaxy flyby simulations can be run with these 6 parameters
specified. We use physical values for all the quantities to facilitate such idealized setups.
We track the orbit of the secondary halo during the flyby and mark the minimum separation between 
the primary and secondary centers as $b$. We also note the primary half-mass
radius (determined from the particle distribution) at the time of this minimum separation.
However, since $b$ is the minimum separation found during snapshot outputs, it is really an upper limit
on the true minimum separation between the two halos. To better estimate the true impact parameter,
we numerically integrated the orbit of two analytic dark matter halos with the same initial total mass,
separation, and velocity as our target halos. To simplify, we assumed the halo concentrations were given by~\citep{BK01}.
As we integrate the orbital trajectory, we include the effect of tidal stripping to the Roche lobe, as well as 
Chandrasekhar dynamical friction with a Coulomb logarithm that goes as log$(1+(M_{\rm prim}/M_{\rm sec})^2)$.
In practice, the mean impact parameter from the simulations was only 10\% times larger than from integrating the orbit, but there was
a non-negligible fraction of impact parameters that were larger by order unity -- these encounters were typically
highly radial and the likelihood of the simulation snapshot catching it at its true pericenter is small. 
We will use the  $b$ determined from this extrapolated trajectory for the rest of the paper.

\subsection{Perturbations from Flybys}\label{subsection:flyby_perturbations}

As a first estimate of the potential damage done by flybys, we follow the prescription set out by \citet{VW00}. Their work is based on linear 
perturbation theory to 
determine the change in total and potential energy for spherical systems. The perturbation equations have different forms depending on whether the halo 
experiences an internal or external encounter (see Appendix~\ref{appendix:perturbations}). 
The generic flyby consists of a smaller secondary halo falling into a more massive primary halo and then detaching at some later
time. In the reference frame of the primary, it is a victim of an internal flyby; for the secondary halo, however, the 
center of the primary halo always remains outside $\Rvir$ so it operates as an external flyby. Therefore, we use the external and 
internal flyby perturbation from \citet{VW00} to calculate the perturbations on the secondary and primary halo respectively. 
We combined the internal and external flyby perturbations fits into one equation: 
\begin{equation}
\label{eqn:combined_perturbation}
\frac{\Delta E_1}{E_1} = \left(\frac{M_2}{0.1 M_1}\right)^2 \exp(-0.5\beta) \exp(\beta  b') {\Vsep^{-\alpha}}, 
\end{equation}
where we replaced $K$ with $\exp(-0.5\beta) \times \exp(\beta b')$, and we used $\alpha=1.95, \,\beta=-1.0$. This provides a continuous fit 
for \dphi in Eqn.~\ref{eqn:internal_perturbation} while preserving the energy changes from the two equations. Essentially, we used an exponential cut-off
from the internal flyby perturbation calculation and extended the prescription to $b > 2\, \var{R}{half,primary}$. Since this is 
somewhat of an ad-hoc implementation, the exact perturbation will not be accurate; however, in this paper we are 
only concerned with relative estimates and these equations should be sufficient for our purposes. In contrast, \citet{LHS14} performed high-resolution 
simulations of galaxy flybys and directly measured the perturbed modes and their magnitudes; they found that the 0.1 mass ratio secondary during a deep flyby could excite a perturbation 
in the $m=2$ mode as high as 20\% (with larger perturbation amplitudes for lower-order modes), resulting in a distinct and long-lived bar.

\section{Results}\label{sec:results}
In \citet{SH12}, we showed that flybys occur frequently for massive halos ($ \gtrsim 10^{11}\,\hinvMsun$) for $z\lesssim 2$. Here, we
show that flybys are potentially an important player in galaxy evolution. 

First, we compare the frequency of flybys vs mergers from a more observational perspective.
Table~\ref{table:flyby_fractions} lists the fraction of flybys and mergers for various host halos at a range of redshifts and relative velocities. 
It is striking that for $z < 2$, the potential flyby contamination in a galaxy-pair survey is as much as $\sim 85\%$, and at minimum, the contamination 
is roughly 1/3. It appears that galaxy surveys designed to probe the effects of mergers on star formation, AGN activity, or morphology could be 
picking up a significant fraction of flybys as well, particularly for Milky Way-mass halos. In this case, flybys do `matter' because they are an 
abundant contaminant.

\begin{longtable*}[ht]{cccccc}
\caption{This table lists the number of flybys and mergers for various redshift, primary mass, and 
different velocity offsets for galaxy pairs. Column 1 shows the redshift range for the beginning of
the interaction, column 2 shows the virial velocity of the primary halo, column 3 shows the relative
physical velocity of the secondary halo (i.e., a velocity offset), column 4 and column 5 show the number of 
flybys and mergers respectively while column 6 shows the percentage of flybys ( = ${Column\, 4}/({Column \,4 + Column\, 5})$). 
The mass-ranges corresponding to Column 2 are $10^{11}-10^{12}\,\hinvMsun$ and $10^{12}-4\times10^{13}\,\hinvMsun$.  
In principle, this table shows the fraction of interactions that will be flybys from an initial halo pair with
a given primary virial velocity and a velocity offset. For $z=1-2$, the flyby rate is {\em at least} comparable to
mergers for all halo masses. }
\label{table:flyby_fractions} \\

\hline\hline \hline \\[1ex]
   \multicolumn{1}{c}{\textbf{redshift range}} &
   \multicolumn{1}{c}{\textbf{Primary \Vvir}} &
   \multicolumn{1}{c}{{$\mathbf{V_{\rm infall}}$}} &
   \multicolumn{1}{c}{\textbf{nFlybys }} & 
   \multicolumn{1}{c}{\textbf{nMergers}} &
   \multicolumn{1}{c}{\textbf{Flyby fraction}}\\[1.0ex]
   \multicolumn{1}{c}{[\textbf{--}]} &
   \multicolumn{1}{c}{[\textbf{km/s}]} & 
   \multicolumn{1}{c}{[\textbf{km/s}]} &
   \multicolumn{1}{c}{[\textbf{--}]}   &
   \multicolumn{1}{c}{[\textbf{--}]} &
   \multicolumn{1}{c}{[\textbf{\%}]} \\[1ex]\hline \hline \\
\endfirsthead

\multicolumn{6}{c}{{\tablename} \thetable{} -- Continued}\\[0.5ex]
  \hline\hline \hline \\[1ex]
   \multicolumn{1}{c}{\textbf{redshift range}} &
   \multicolumn{1}{c}{\textbf{Primary \Vvir}} &
   \multicolumn{1}{c}{{$\mathbf{V_{\rm infall}}$}} &
   \multicolumn{1}{c}{\textbf{nFlybys }} & 
   \multicolumn{1}{c}{\textbf{nMergers}} &
   \multicolumn{1}{c}{\textbf{Flyby fraction}}\\[1.0ex]
   \multicolumn{1}{c}{[\textbf{--}]} &
   \multicolumn{1}{c}{[\textbf{km/s}]} & 
   \multicolumn{1}{c}{[\textbf{km/s}]} &
   \multicolumn{1}{c}{[\textbf{--}]}   &
   \multicolumn{1}{c}{[\textbf{--}]} &
   \multicolumn{1}{c}{[\textbf{\%}]} \\[1ex]\hline \hline \\
\endhead

\multicolumn{3}{l}{{Continued on Next Page\ldots}} \\[1ex]
\endfoot

  \\[-2ex] \hline\hline\hline
\endlastfoot

\multirow{5}{*}{{$1.0-2.0$}}
                           & \multirow{3}{*}{{85--190}}   &   170--380    &       1789  &         1082   &         62  \\
                           &                              &   255--570    & \phno  200  & \phnt     66   &         75  \\
                           &                              &   340--760    & \phnt   32  & \phno\phnt 5   &         86  \\[0.8ex]
                           & \multirow{2}{*}{{190--645}}  &   380--1290   & \phno  911  &         1256   &         42  \\
                           &                              &   570--1935   & \phnt   29  &  \phnt    60   &         33  \\[1.5ex]\hline\\
\multirow{4}{*}{{$2.0-3.0$}}
                           & \multirow{2}{*}{{100--220}}  &   200--440    &        692  &         1022   &         40 \\
                           &                              &   300--660    & \phno   37  &  \phnt    68   &         35 \\[0.8ex]
                           & \multirow{2}{*}{{220--750}}  &   440--1500   &        176  &  \phno   567   &         24 \\
                           &                              &   660--2250   & \phnt    9  &  \phnt    72   &         11 \\[1.5ex]\hline\\
\multirow{4}{*}{{$3.0-4.0$}} 
                           & \multirow{2}{*}{{115--250}}  &   230--500    &        155  &          541   &         22 \\
                           &                              &   345--750    & \phnt    3  & \phno     19   &         14 \\[0.8ex]
                           & \multirow{2}{*}{{250--845}}  &   500--1690   & \phnt    7  &          112   &    \phno 6 \\
                           &                              &   750--2535   & \phnt    0  & \phno     15   &    \phno 0 \\[1.5ex]
\end{longtable*}

\subsection{Two classes of flybys}
A flyby will be perturbative when the halos penetrate
deeply before detaching; in addition, if the flyby is relatively slow, then the secondary halo particles spend more time near the resonances 
-- thereby perturbing the primary even more. Thus, highly perturbative flybys would have to be highly penetrating and relatively slow. In
the next sections, we will dissect the orbital characteristics of the flybys in the simulation and examine their relative importance
over a range of halo masses and redshift. 

\begin{figure}
\centering
\includegraphics[width=\linewidth,clip=true]{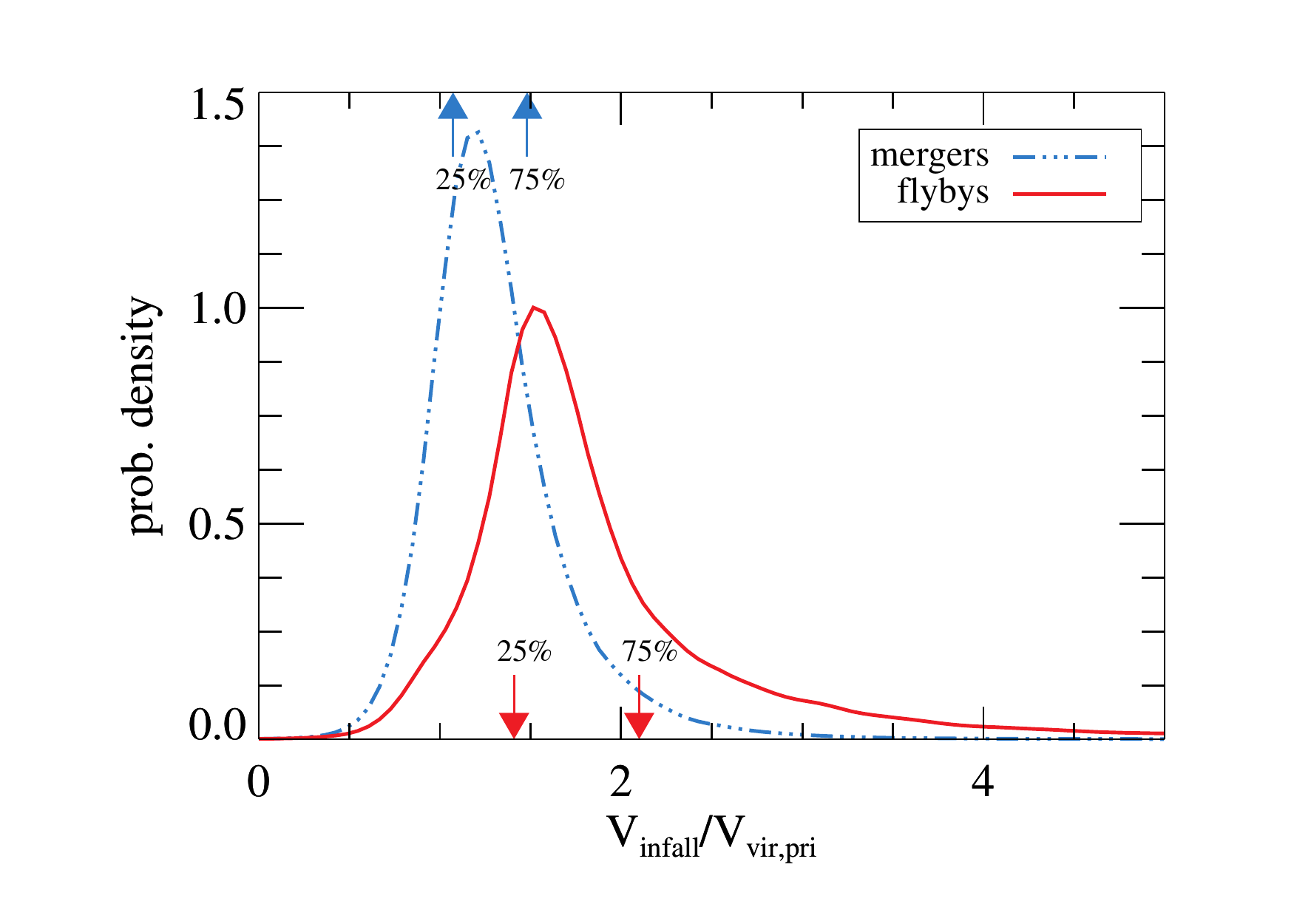}
\caption{\small Probability density of infall velocities (scaled by the primary \Vvir) for mergers (blue, dash-dotted line)
and flybys (red, solid line). The merger infall velocity peaks at $\sim 1.2 \Vvir$ while the flyby infall velocity peaks 
at a higher value of $\sim 1.6\Vvir$. Formally, the escape speed for the interaction is $\sqrt{2}\Vvir$. Hence, most of 
the flybys have infall velocities much larger than the escape speed of the primary halo. A two-sided K-S test shows that 
merger and flyby infall velocity distribution come from different populations (p-value for the null hypothesis that they 
are the same distribution is $< 10^{-8}$).}
\label{fig:ratio_vsep_to_vvir_flybys_and_mergers}
\end{figure}

First, we identify the infall characteristics of flybys. For a flyby to occur, the 
secondary halo has to overcome the primary halo's gravitational potential; therefore, the initial relative velocity
must be high compared to the primary halo escape velocity. 
A priori, therefore,  we expect that flybys will have an higher initial relative velocity compared to mergers. 
Nominally, an unbound interaction should have $\vsep \geq {V_{esc,primary}} = \sqrt{2} \Vvir$; however, 
mass-loss and dynamical friction effectively increase this minimum infall velocity.  
Previous work looking at the distribution of 
subhalo infall velocities has found that typical $v_{\rm infall} \sim \Vvir$ and only a few percent of the orbits are truly unbound~\citep[e.g.][]{B05,LNSJFH09}
In Fig.~\ref{fig:ratio_vsep_to_vvir_flybys_and_mergers} we plot the histogram of the relative velocity scaled by the primary $\Vvir$ for both mergers and flybys.
We note that flybys and mergers are indistinguishable in a plot for 
the ratio of $\rsep$ to $\Rvir$. Since we classify flybys based on past/future behavior depending on the velocities of the halos, 
there is no distinction between mergers and `grazing' flybys in real space. As expected, the infall velocity of mergers peaks at a lower (bound) velocity, $\sim 1.2 \Vvir$  compared to the peak flyby infall 
velocity of $\sim 1.6\,\Vvir$. We find that roughly 1\% and 10\% of the flybys occur with $\vsep \le \sqrt{2}$ and $2$  times the 
primary $\Vvir$ respectively. For secondary halos, about 50\% of the flybys
occur at relative velocities $> 5\times \Vvir$, while only $\sim$ 30\% of the mergers occur at $\vsep \gtrsim 5\times \Vvir$.

Fig.~\ref{fig:ratio_vsep_to_vvir_flybys_and_mergers} demonstrates that flybys have a higher infall velocity compared to mergers
but the initial separation between centers for both mergers and flybys is approximately the sum of the virial radii of the two halos. 
So, the only way to observationally distinguish between a flyby and a merger would be through the 3-d relative velocity,
which makes it problematic to disentangle mergers from flybys in a survey.

\begin{figure}
\centering
\includegraphics[width=0.8\linewidth,clip=true]{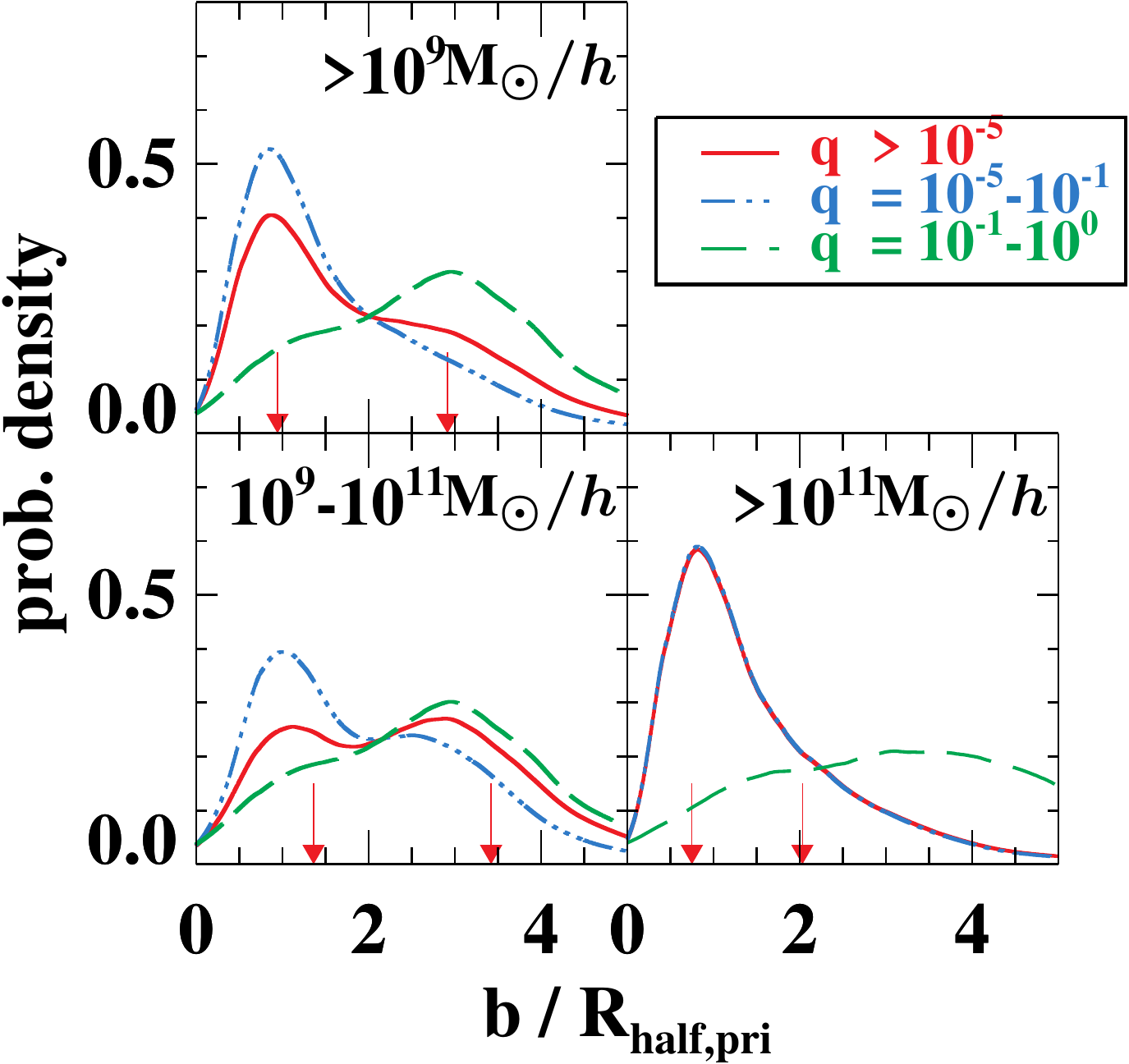}
\caption{\small The probability density of flyby impact parameters for different mass-ratios and
primary masses. The top left panel shows all primary masses  $ > 10^9\,\hinvMsun$,
the bottom left shows $10^{9}-10^{11}\,\hinvMsun$ while the bottom right panel is for all primary
masses above $> 10^{11}\,\hinvMsun$. The (red) solid lines shows all flybys with $q>10^{-4}$,  
the (blue) dash-dotted line shows $q=10^{-4}-10^{-1}$, while the (golden yellow) dashed line shows
major flybys with $q=10^{-1}-10^0$. The two arrows show the 25\% and 75\% quartile for all the 
flybys (i.e., $q>10^{-4}$) in that panel. 
From the plots, we can see that the typical $b$ for a flyby is $\sim$ primary $\Rhalf$ for mass-ratio $q \lesssim 10^{-1}$.
The flybys with $q>0.1$, almost inevitably are distant. We postulate that this is an outcome of efficient dynamical
friction for close-to equal-mass interactions, where the subhalo loses its kinetic energy and cannot escape
the main halo potential for deep orbits. Therefore, a typical `major' flybys occurs with $b \gtrsim 3\,\Rhalf$ 
and is essentially a grazing encounter. }    
\label{fig:flyby_impact_params_with_q}
\end{figure}

We can also characterize the distribution of the flyby impact parameter. In Fig.~\ref{fig:flyby_impact_params_with_q},
we show the probability density of $b$, scaled by primary $\Rhalf$ at the time of minimum separation, for three
primary halo mass ranges and three mass-ratio cuts. We calculate the probability density with an adaptive
kernel density estimator~\citep{S98} with a Epanechnikov kernel. Combining the impact parameter results with the 
flyby velocity distribution above, it is clear that 
there are two classes of flybys: one that rapidly delves into the primary core, and one that slowly skirts the outer primary halo. 
For massive halos, the most common flyby is a deep one, with few skirting encounters. 
In general, major flybys are of the slow and shallow type, while minor flybys penetrate deeper into the primary halo potential.

\subsection{How long do typical flybys last?}
We have so far looked at the distribution of $b$ and \vinfall for flybys. Now, another important orbital characteristic
of flybys will be the typical duration. We remind the reader that a flyby is characterized by a 
`main halo$\rightarrow$subhalo$\rightarrow$main halo' transition, and
dynamically this means that the secondary halo must fall in to the primary halo and climb out of the potential well. 
This implies that the typical flyby duration is $\mathcal{O}(\var{t}{cross})$.   
In fact, in \citet{SH12}, we used a minimum duration of half a crossing time to distinguish `true' flybys 
from artificial halo stitching due to the \FOF algorithm. We choose the crossing time at the beginning of the 
flyby, \var{t}{cross}, to express the flyby durations. By selection, all flybys must last {\em at least} $0.5$ \var{t}{cross}. 
We note that \var{t}{cross} is independent of halo mass and evolves as $H(z)^{-1}$ with $z$. 

\begin{figure}
\centering
\includegraphics[width=\linewidth,clip=true]{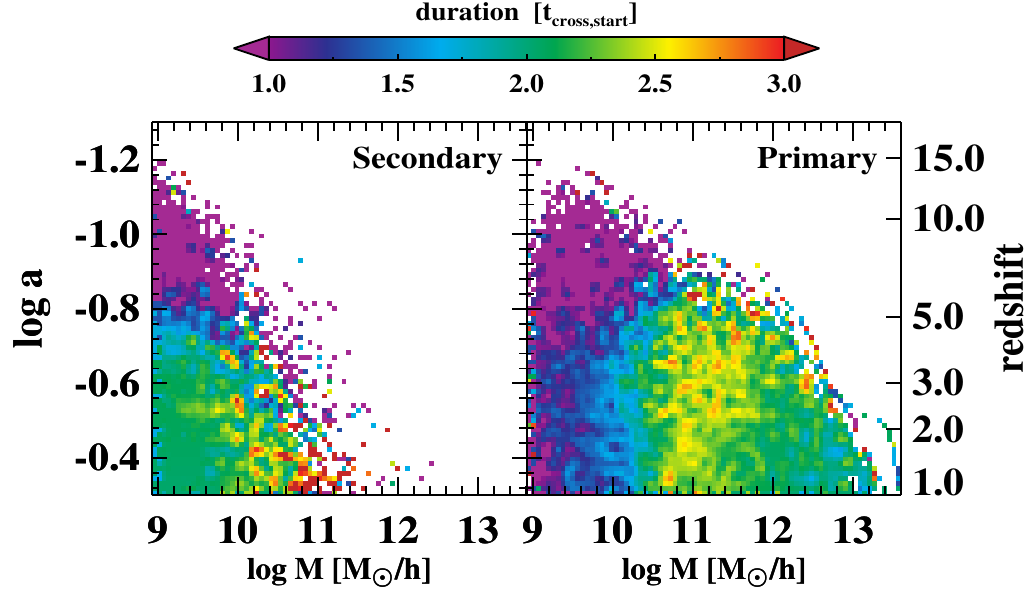}
\caption{\small Median flyby duration, in units of \var{t}{cross} as a function of
halo mass and redshift. Primary halos are plotted on the right and secondary halos are plotted on the 
left. The redshift and halo mass used for the plot is at the beginning of the flyby. Yellow/red means that 
for flybys in that mass-$z$ bin, the median flyby will last $\gtrsim \, 2-3$ \var{t}{cross} for that redshift. 
At $z\gtrsim8$, flybys are very fast and typically last one \var{t}{cross} ($\sim \, 50-70$ Myrs)
whereas at lower redshift, $z\lesssim5$, flybys typically last $2-3$ \var{t}{cross} ($\sim\, 1-2$ Gyr). }
\label{fig:flyby_duration}
\end{figure}

In Fig.~\ref{fig:flyby_duration} we show the median flyby duration as a function of halo mass and redshift for both the
primary and secondary halos. The duration is scaled by \var{t}{cross} -- where \var{t}{cross} is evaluated at the beginning 
of the flyby. From the figure, we can see that the median duration of a flyby is typically larger than two \var{t}{cross}.
There is an redshift trend for the duration as well -- high $z$ flybys are shorter compared to the low $z$ ones. For $z\lesssim 5$,
both primary and secondary halos typically undergo flybys that last 2--3 \var{t}{cross}. However, since we can only detect 
the beginning and end of flybys at a snapshot (which goes as log$a$), we caution that the true flyby duration may be shorter.

\subsection{Distribution of mass ratios}\label{subsection:massratios}
\begin{figure}
\centering
\includegraphics[width=\linewidth,clip=true]{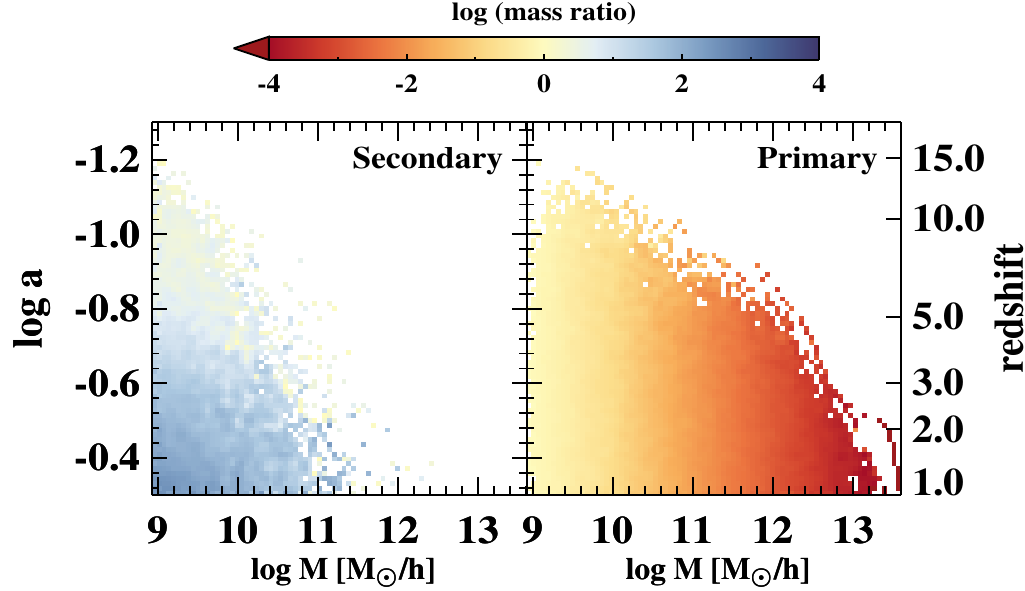}
\caption{\small Distribution of the median mass ratio of flybys as a function of halo mass 
and redshift. Primary halos are plotted on the right while secondary halos are 
plotted on the left. Our formal halo detection limit is $\sim 10^8\,\hinvMsun$ but we 
only use halos with at least 100 particles, i.e., $\sim 5\times\,\hinvMsun$ halos. 
Thus, for the low mass primary halos, e.g., $\sim 10^9 \hinvMsun$, all interactions are
resolution limited and appear roughly equal mass for all redshifts. In the same spirit,
the typical interaction for a massive halo is with a $\sim 10^9\,\hinvMsun$ and the 
smallest mass ratio can be as low as $\sim 10^{-4}$. For a massive secondary halo, there 
are very few more massive halos and consequently, the typical interaction (if there is one),
is more equal-mass. }
\label{fig:mass_ratio}
\end{figure}

In any cosmological simulation, the halo number density is highest for the lowest resolvable halo. Consequently, the frequency of interactions increase 
with decreasing mass ratio, roughly as $q^{\massratioindex}$. In Fig.~\ref{fig:mass_ratio}, we show the distribution of median flyby mass ratios for primary 
and secondary halos. For primary halos, the trend is roughly independent of redshift, in other words, the mass ratio smoothly decreases with increasing halo 
mass. The lowest mass primary halos have flybys that are approximately equal-mass, while the typical flyby for the most massive primary halos is
a $q \sim 10^{-4}$. \footnote{ The secondary halos are less massive than the primary halos by construction -- therefore the mass ratio for the secondary halos 
are always greater than 1.} 

Note that interpreting this figure is not straightforward because it also encodes the halo mass function. 
For example, for the massive secondary halos, e.g, $10^{11}\,\hinvMsun$ at $z\sim 5$, the only feasible primary halos are also 
of similar mass. Hence, the mass ratio for such flybys is close to 1. For a low-mass secondary halo, the typical interaction is $q \sim 10^{4}$ --
the converse of its high-mass primary halo counterpart. 
Here we see that flybys very rarely occur among massive, equal-mass halos. While one reason is that any major interaction
is always less likely compared to a minor one, we do still see major mergers in the simulation. So, naturally the question
arises -- why are there fewer major flybys? The answer comes from the velocity function of dark matter subhalos. In \LCDM cosmology, the velocity
of infalling satellites drops of steeply, as $\exp(- (v_{sat}/v_{host})^\alpha)$, with $\alpha \sim 3$~\citep{HW06}. In addition, the efficiency of
dynamical friction and tidal stripping ensures that massive subhalos get stripped, lose significant amount of energy and angular momentum, and have 
a more difficult time making it out of the primary halo. This effect was pointed out by \citet{LNSJFH09}, who noticed the mass-bias in unorthodox orbits -- 
where smaller subhalos can preferentially lie as far as $5\,\Rvir$ from the primary host halo.    The initial velocity distribution and the preferential 
energy loss for massive halos combine to make major flybys comparatively rare.

\subsection{Mass loss from flybys}

\begin{figure}
\centering
\includegraphics[width=\linewidth,clip=true]{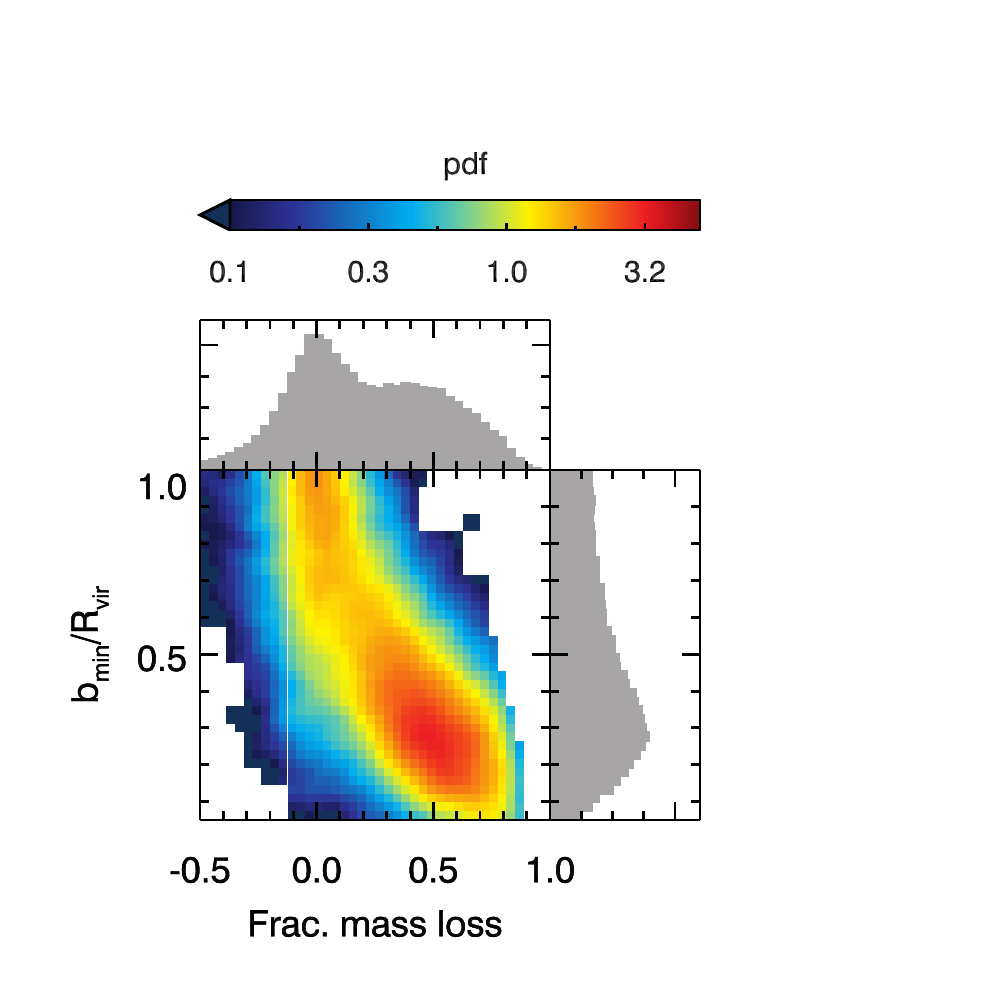}
\caption{\small Mass loss of the secondary ($1-M_{final}/M_{infall}$) as a function of the impact parameter (scaled by the host \Rvir). 
There are two distinct peaks in the distribution: i) deep flybys and ii) grazing flybys. Grazing flybys typically do not
lose much of their halo mass while the deep flybys can lose $\sim 50\%$ of the infall halo mass.}
\label{fig:flyby_massloss}
\end{figure}

Now that we have discussed the demographics of flyby encounters, we can turn our attention to the effect that flybys exact on
the participating halos. This can address the question of whether flybys matter in transforming the halo structure and kinematics.
One key effect is that the interaction can cause significant mass loss in the intruder halo. This is quantified in
Fig.~\ref{fig:flyby_massloss}, which plots the fractional mass loss of the secondary halo as a function of the normalized impact parameter.
This draws out two essential points: first, deep encounters strip the secondary halo mass by of order 50\%, and second, the grazing encounters may 
sometimes gain mass as they travel along the equipotential contour of the primary halo.
From this result, we can already answer our original question of {\em do flybys matter?} with a definitive yes, 
because they dramatically alter the mass (and therefore the mass-to-light ratio) of the intruder halo.

\subsection{Flyby-driven perturbations}

Taking the concept of flyby-driven halo transformation a step further, we quantify the frequency of perturbative flybys. Broadly speaking, the perturbation
from a flyby depends on the mass ratio, the impact parameter and the relative velocity. The usual intuitive understanding of interactions apply --
more massive secondaries cause more damage, as do smaller impact parameters and lower relative velocities.  

\begin{figure}
\centering
\includegraphics[width=\linewidth,clip=true]{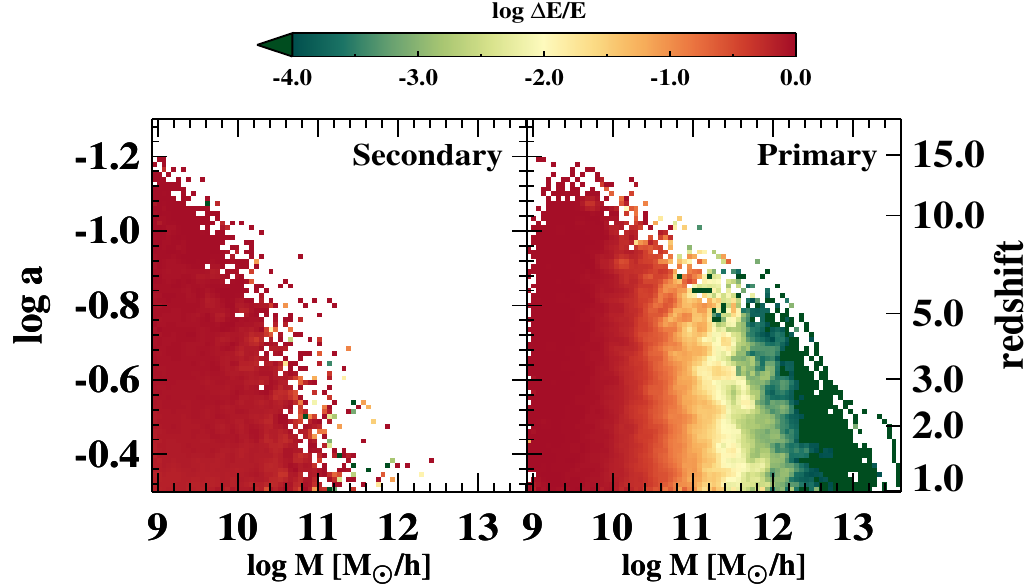}
\caption{\small The mean of \dphi as a function of halo mass and redshift. The primary mass is 
plotted on the right and the secondary mass on the left. We show the mean (and not the median) here to 
have a more representative measure of the perturbation on the halos. It is clear that the mean perturbation
from flybys is much more damaging for the intruder halos than the primary halos. }
\label{fig:flyby_mean_deltaE}
\end{figure}

With all the essential pieces now in hand, we will look at the perturbations from flybys -- these perturbations 
are estimated using Eqn.~\ref{eqn:combined_perturbation}. 
Looking at the functional form of Eqn.~\ref{eqn:combined_perturbation}, the most important bit is the impact parameter, $b$ which 
determines if  \dphi is characterized as an internal or external flyby. In Fig.~\ref{fig:flyby_mean_deltaE}, we show the 
mean perturbation in 2-d cells of halo mass and redshift. For the secondary halos (left panel), we see that the mean \dphi 
during a flyby is of order unity and should strongly affect the internal structure of the secondary halo\footnote{\dphi of 
order unity implies the assumed linear regime for deriving Eqn.~\ref{eqn:combined_perturbation} is no longer valid}. However, for the 
primary halos (right panel), the mean \dphi from a flyby is essentially negligeble. Thus, for massive primary halos, a 
typical flyby is not going to change the overall structure at all. 

\begin{figure}
\centering
\includegraphics[width=\linewidth,clip=true]{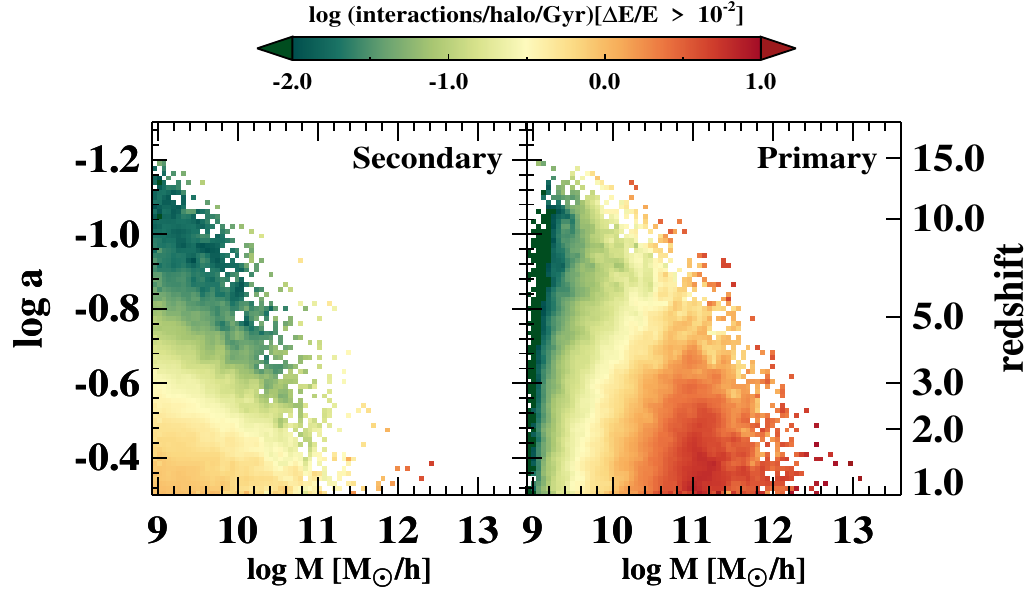}
\caption{\small The rate of big flybys ($\dphi \gtrsim 10^{-2}$) as a function of halo mass and redshift. The rate 
is plotted on the right for primary halos and on the left for secondary halos. The color shows the logarithm of 
the flyby rate per halo per Gyr. Even though in Fig.~\ref{fig:flyby_mean_deltaE} we saw that the mean \dphi 
is small for massive primary halos, these massive halos undergo so many flybys that there is {\em always}
some big perturbative flyby that occurs at any given $z<3$. For instance, all halos $\gtrsim 10^{11}\,\hinvMsun$ at
$z<3$, have a perturbative flyby rate $\gtrsim 1$ per Gyr. Such frequent disturbances will influence the evolution 
of all such massive halos.} 
\label{fig:big_flyby_rate}
\end{figure}

However, since the big question we want to know is if {\em flybys matter}, we need to know if {\em any}, and not just the typical flyby 
strongly affects a halo/galaxy. In Fig.~\ref{fig:big_flyby_rate}, we show the event rate of perturbative flybys, defined as $\dphi > 10^{-2}$, as a function of 
halo mass and redshift. On the left panel, we show the event rate for `big' flybys for secondary halos. The average event rate for secondaries 
increases with decreasing redshift; with typical rates $\sim 0.01$ per Gyr at $z \gtrsim 4$, and $\sim 0.3$ per Gyr for $ 1 \gtrsim z \gtrsim 4$. 
The event rate resembles the distribution of durations (see Fig.~\ref{fig:flyby_duration}, and shows almost no trend with halo mass. The similarity 
with the duration is likely from the fact that deeper flybys take longer to complete, and these deeper flybys produce larger perturbations. For the 
primary halos, there is a strong trend with halo mass -- with the event rate increasing by almost an order of magnitude between $10^{10} \hinvMsun$ and 
$10^{11} \hinvMsun$ halos. All halos above $10^{11} \hinvMsun$ have {\em at least} a few `big' flybys every Gyr. Thus, both primary and secondary 
halos have perturbative flybys frequently enough that the halo internal structure will be modified~\citep{LHS14,KPKAY14}.

In Fig.~\ref{fig:flyby_mean_deltaE}, we saw that the typical flyby was highly perturbative for all secondary halos but in Fig.~\ref{fig:big_flyby_rate}, 
we see that the typical secondary halo only experiences $\sim 1$ highly perturbative flyby per Gyr. Similarly, in Fig.~\ref{fig:flyby_mean_deltaE}, 
we saw that the typical flybys don't significantly perturb the primary, whereas in Fig.~\ref{fig:big_flyby_rate}, we see that 
primary halos undergo a large number (few to 10's) of highly perturbative flybys per Gyr. This apparent discrepancy between 
Fig.~\ref{fig:flyby_mean_deltaE} and Fig.~\ref{fig:big_flyby_rate}, can be explained in terms of the hierarchical nature of halo interactions, 
and the perturbation calculation (Eqn.~\ref{eqn:combined_perturbation}) itself. Secondaries 
don't always have flybys, but when they do, it is a highly perturbative one. Averaged over the entire population of secondary halos, the event 
rate of highly perturbative flybys is small. On the other hand, massive primary halos undergo a lot of flybys; however, most of these are with 
small secondaries and the $q^2$ term in Eqn.~\ref{eqn:combined_perturbation} reduces the typical \dphi drastically. Thus, the typical interaction 
strength for the primary halo is actually very small. But, in this large number of tiny flybys, there are a few interactions where the secondary gets 
close to the center of the primary halo. These flybys, with small impact parameters, produce really large \dphi and drive the high event rate 
for massive primaries.

\section{Discussion}\label{sec:discussion}

Flybys and mergers have similar separation distributions, differing only by the relative velocities between the halos.
One obvious implication of this similarity in separation is that it is very difficult to distinguish flybys from mergers in a photometric survey, and even
adding radial velocities will only help in a statistical sense. We should expect that merger
rates derived from pair counts will be contaminated with flybys. However, since
flybys typically occur with low mass-ratios, this contamination affects minor merger rates much more
than major merger rates.

One consistent picture that emerges from flybys is that the secondary halos typically lose $\sim 50\%$ 
of their infall mass, and are highly perturbed as well. Previous work has shown that this
level of perturbation drives bars and warps, and spins up the halo. Another potential side-effect of a flyby could be
the shut-off in smooth gas accretion and consequent quenching of star formation in the
secondary as it delves into the primary halo. 

In addition to these potential changes to internal galactic structure,
flybys also have an effect on the overall galaxy distribution.  Flyby-induced mass loss,
while keeping the baryonic component of the secondary galaxy relatively intact,
systematically decreases its mass-to-light ratio (\ml) at fixed luminosity. 
Observationally, these would imply a higher mass-to-light ratio (\ml) and redder colors 
for a fraction of galaxies that are near other more massive galaxies. Since flybys are hierarchical, i.e.,
more massive halos have more flybys~\citep{SH12}, the effect of flybys would be more prominent
in galaxy groups and clusters. Such an excess in the red-fraction has been reported~\citep{WYMBK09}, and more
recently \citet{WTCB13} showed that that the excess in red-fraction could be explained by assuming
flyby galaxies evolved like satellites even outside the host.

Another effect of tidal stripping during a flyby shows up as {\it halo assembly bias}. Halo assembly bias means
that the clustering of halos depends on a hidden parameter beyond the halo mass.
Over the past decade, halo assembly bias has been consistently detected in simulations~\citep[e.g.,][]{WZBKA06,GW07,WMJ07,LP11} and
some have invoked `ejected' subhalos as an explanation. However, a far more important consequence
of flybys would be {\it galaxy} assembly bias -- where galaxy properties depend not just on the host
halo mass but another secondary characteristic, such as environment. If flybys systematically reduce halo masses, a
galaxy-mapping prescription based only on halo mass would erroneously place a 
lower mass, bluer galaxy within. Or, worse still, the halo may not even receive a galaxy. Comparing this mock galaxy catalog with observations, the mock would show a dearth of red galaxies around massive systems,
as well as a suppression of galaxy clustering. As emphasized recently~\citep{ZHB13}, galaxy assembly bias can cause Halo Occupation Distribution (HOD) models to
converge on a statistically different parameter space. 
Galaxy assembly bias from flybys would require an overhaul of the models that place galaxies into halos and statistically model 
galaxy formation physics, such as the HOD~\citep{BW02}. Similar concerns apply for the Conditional Luminosity Function (CLF)~\citep[e.g.,][]{YMB03}, as well.

\section{Conclusions}\label{sec:conclusion}
In this paper, we characterized halo flybys by their infall velocity, impact parameter, duration, mass-ratio distribution, 
as well as the mass loss and perturbation due to the flyby on both halos. This will be useful for any future numerical or semi-analytic project
that wishes to examine the effect of flybys using realistic distributions. We summarize our main results below:
\begin{enumerate}
\item Flybys occur with higher \var{v}{infall} velocities compared to mergers. The mean infall velocity for flybys and mergers are $\sim 1.8$ and $1.3$ the virial velocity of the primary halo.
respectively. See Fig.~\ref{fig:ratio_vsep_to_vvir_flybys_and_mergers} for the probability density of the infall velocities. 
\item Flybys typically penetrate to about the half mass radius of the primary halo, but can go as deep as $\lesssim$ 0.01 \Rvir.
However, the impact parameter is strongly dependent on the mass ratio; more minor
flybys usually have smaller impact parameter. See Fig.~\ref{fig:flyby_impact_params_with_q} for the probability density of flyby impact parameters. 
\item There are two classes of flybys: slow interactions that skirt the outer halo, and fast flybys that delve into the core.
\item Flybys typically last roughly $4-5$ \var{t}{cross} for $z\lesssim 5$. There is a slight mass dependence -- less massive primary halos under-go shorter-lived
flybys. At high $z$, most flybys are quite short in units of crossing time.  See Fig.~\ref{fig:flyby_duration} for the distribution of flyby durations.
\item Flybys and mergers have a similar dependence on mass-ratio, $q^{\massratioindex}$. Major flybys are very rare (even more so than major mergers); the largest $q$ we found was $\sim 0.4$ in our small volume. However, at high $z$, flybys occur between similar mass halos. At lower $z$, massive primary halos undergo 
flybys with small halos, with a typical $q \sim 10^{-3}$. See Fig.~\ref{fig:mass_ratio}. 
\item Deep encounters strip about 50\% of the secondary halo mass, while grazing encounters exhibit negligible mass loss (and may even cause a gain in secondary mass). See Fig.~\ref{fig:flyby_massloss}. 
\item We find that flybys are usually highly perturbative for secondary halos (of order unity perturbation) but are typically extremely weak for the primary halos ($\sim 10^{-4}$, see Fig.~\ref{fig:flyby_mean_deltaE}). Since the 
mass-ratio for a typical flyby, is $\sim 10^{-3}$, the perturbation, $\dphi \propto q^2$, is tiny. However, since massive primary halos undergo a large number of flybys, some of these flybys are highly perturbative. In 
Fig.~\ref{fig:big_flyby_rate}, we show that highly perturbative flybys occur at $\gtrsim 1$ per Gyr for all halos $\gtrsim 10^{10} \hinvMsun$ for $z \lesssim 4$.\footnote{So, contrary to standard lore, the answer 
to the question in the paper-title is not always `no'}
\end{enumerate}

\acknowledgments
This work was conducted in part using the resources of the Advanced Computing Center for Research and Education at Vanderbilt University, Nashville, TN. 
We also acknowledge support from the NSF Career award AST-0847696.

\bibliographystyle{apj}
\bibliography{Biblio-Database}

\begin{appendix}
\section{Calculating the perturbations from flybys}\label{appendix:perturbations}
Flybys are high-speed encounters between halos -- therefore, one could assume that the impulse approximation is valid. 
Flybys will impart an impulse on the secondary halos, resulting in an initial increase in the kinetic energy of
the secondary. If the secondary was in equilibrium prior to the flyby, this impulse will throw it out of equilibrium and cause
it to expand. Under the impulse approximation, the change in energy of the secondary~\citep{S58,AW85} is:
\begin{equation}
\Delta E_2 = \frac{4}{3} G^2 M_2 \left(\frac{M_1}{\vsep}\right)^2 \frac{{\rsep}^2}{b^4}.
\end{equation}\label{eqn:impulse}
For an extended perturber, the right hand side of Eqn.~\ref{eqn:impulse} is modified with an additional factor~\citep{GLO99a}, $f(b)$,  where
$f(b)$ changes from 1 for $b=\infty$ to 0 for $b \approx 0$. In that modified form, the largest $\Delta E_2$ occurs for $r_{\rm sep} \sim {R_1}$.
The `grazing' flybys that we study approximately have $r_{\rm sep} \sim R_{\rm vir,1}$ and should thus have the strongest
perturbations. However, both the impulse and tidal approximation disregard resonant effects in the secondary. 
To get a more accurate estimate of the overall change in the potential due to a flyby, we use the empirical results in Table 3 from \citet{VW00}.
Briefly, the change in energy of a perturbed system is given by:
\begin{equation}
\label{eqn:external_perturbation}
\frac{\Delta E_2}{E_2} = \frac{{M_1}^2}{{M_2}^2} \frac{{\rm R_{vir,2}}^6}{{\Rsep}^6} \times 
\begin{cases}
1.3\times10^{-3}   \qquad          &\Omega \leq  0.5, \\
7.0\times10^{-4}   \qquad   0.5 <  &\Omega \leq  1.0, \\
4.7\times10^{-4}   \qquad   1.0 <  &\Omega \leq  2.5, \\
2.8\times10^{-4}   \qquad   2.5 <  &\Omega \leq  5.0, \\
0.0                \qquad          &\Omega  >    5.0.\\
\end{cases}
\end{equation}
where, $\Omega = \frac{\Vsep}{\Rsep} \sqrt{\frac{{\rm R_{vir,1}}^3}{G\, M_1}}$, 
$\Rsep$ and $\Vsep$ refer to the relative separation and velocity of the centers of the two halos, $M$ is the total mass of a halo, $G$
is the universal gravitational constant, $\Rvir$ is the virial radius of a halo, $\Phi$ is the total potential energy
of the halo and the subscripts 1 \& 2 refer to the primary (more massive) and the secondary halo respectively. For a given
pair of halos, \dphi is larger for the secondary halo.
For penetrating encounters, we calculated the perturbation, \dphi, of the primary using the fit including the effects of damped 
modes in Table 2 of \citet{VW00}:
\begin{eqnarray}
\frac{\Delta E_1}{E_1} = \left(\frac{M_2}{0.1\times M_1}\right)^2 \times K \times {\Vsep^{-\alpha}} \quad \mbox{where},\\
(K,\alpha)  =
\begin{cases}
(1.0,1.96)   \qquad   & 0 \leq b' <  1, \\
(0.5,1.90)  \qquad   & 1 \leq  b \leq  2.\\
\end{cases}\notag
\label{eqn:internal_perturbation}
\end{eqnarray}
Here, $b'=\Rsep/\Rhalf$, is the impact parameter scaled in terms of the half-mass radius of the primary, and $\Vsep$ is
the relative velocity in units of $200 \, \kms$. However, the internal and external fits are not continuous
functions with impact parameter. We modified the fits to allow for a smooth transition between the internal fit (for
$b \lesssim 2\times\, \mbox{primary} \, \Rhalf$) and the external fit (for $b > 2 \,\mbox{primary}\, \Rhalf$):
\begin{eqnarray}
\label{eqn:appendix:combined_perturbation}
\frac{\Delta E_1}{E_1} = \left(\frac{M_2}{0.1 M_1}\right)^2 \exp(-0.5\beta) \exp(\beta  b) {\Vsep^{-\alpha}}, 
\end{eqnarray}
where, we have replaced $K$ with $\exp(-0.5\beta) \times \exp(\beta b)$. We used $\alpha=1.95, \,\beta=-1.0$. This provides a continuous fit 
for \dphi in Eqn.~\ref{eqn:internal_perturbation} while (roughly) preserving the parameter values from the discrete fit.

\end{appendix}

\end{document}